\documentclass[aps,prl,twocolumn,showpacs,superscriptaddress,floatfix,10pt]{revtex4-1}
\usepackage{framed}
\usepackage{graphicx}
\usepackage{amsmath}
\usepackage{epsfig}
\usepackage{helvet}
\usepackage{amssymb}
\usepackage{framed}
\usepackage[pdftex,plainpages=false,colorlinks=true,linkcolor=blue, citecolor=blue, urlcolor=blue]{hyperref}

\newcommand{\bra}[1]{\mbox{$\langle #1 |$}}
\newcommand{\ket}[1]{\mbox{$| #1 \rangle$}}
\newcommand{\braket}[2]{\mbox{$\langle #1 | #2 \rangle$}}

\newcommand{\frw}[1]{$\overset{\lower0.5em\hbox{$\smash{\scriptscriptstyle\smile}$}} #1$}

\def\B{{\mathcal B}}

\def\L{{\mathcal L}}

\def\IDR{\textrm{\mbox{\tiny IDR}}}
\def\ER{\textrm{\mbox{\tiny ER}}}
\def\K{\textrm{\mbox{\tiny B}}}

\def\MERA{\textrm{\mbox{\tiny MERA}}}

\begin{document}

\title{Implicitly disentangled renormalization}
\author{Glen Evenbly}
\affiliation{D\'epartement de Physique and Institut Quantique, Universit\'e de Sherbrooke, Qu\'ebec, Canada
}
\email{glen.evenbly@usherbrooke.ca}
\date{\today}

\begin{abstract}
We propose a new implementation of real-space renormalization group (RG) transformations for quantum states on a lattice. Key to this approach is the removal of short-ranged entanglement, similar to Vidal's entanglement renormalization (ER) \cite{ER}, which allows a proper RG flow to be achieved. 
However, our proposal only uses operators that act locally within each block, such that the use of disentanglers acting across block boundaries is not required. 
By avoiding the use of disentanglers we argue many tensor network algorithms for studying quantum many-body systems can be significantly improved. The effectiveness of this RG approach is demonstrated through application to the ground state of a $1D$ system at criticality, which is shown to reach a scale-invariant fixed point.
\end{abstract}

\maketitle

\textbf{Introduction.---} The renormalization group (RG) \cite{RGReview} is nowadays regarded as a key conceptual element in both quantum field theory and in condensed matter physics. However the use of RG in \emph{real-space} long proved difficult historically; it was only with the advent of White's density matrix renormalization group (DMRG) algorithm \cite{DMRG1, DMRG2, DMRG3}, which built on the earlier ideas of Kadanoff \cite{Kad1} and Wilson \cite{Wil3}, that real-space RG became firmly established as a powerful computational tool for the study of quantum many-body systems. Vital to the success of DMRG was, in mapping blocks of sites from an initial lattice to effective sites of a coarser lattice, a proper characterisation of the block degrees of freedom that should be retained.

More recently, a significant advance in real-space RG was realized with the proposal of entanglement renormalization (ER) \cite{ER}, which uses unitary \emph{disentanglers} to remove short-ranged entanglement between blocks as part of a coarse-graining scheme. Key advantages of ER include (i) the ability to generate an RG flow with the proper structure of fixed points, such that scale-invariance is realized in quantum critical systems \cite{Crit2, Crit3, Crit4, Crit5}, and (ii) the ability to generate a computationally sustainable RG flow that can be scaled to large system sizes while maintaining a small truncation error \cite{ER,ERalg,ERferm, ERbose}. In addition to the study of quantum many-body systems ER has proven useful in a diverse range of applications including, for instance, for the study of the AdS / CFT duality \cite{Swingle12, Swingle12b}.
However, while conceptually appealing, ER has proven difficult to put to use, especially for systems in $D>1$ dimensions, due to the computational challenges associated with the use of disentanglers \cite{ERalg, 2DMERA1, 2DMERA2}.

\begin{figure}[!t!b]
\begin{center}
\includegraphics[width=6.5cm]{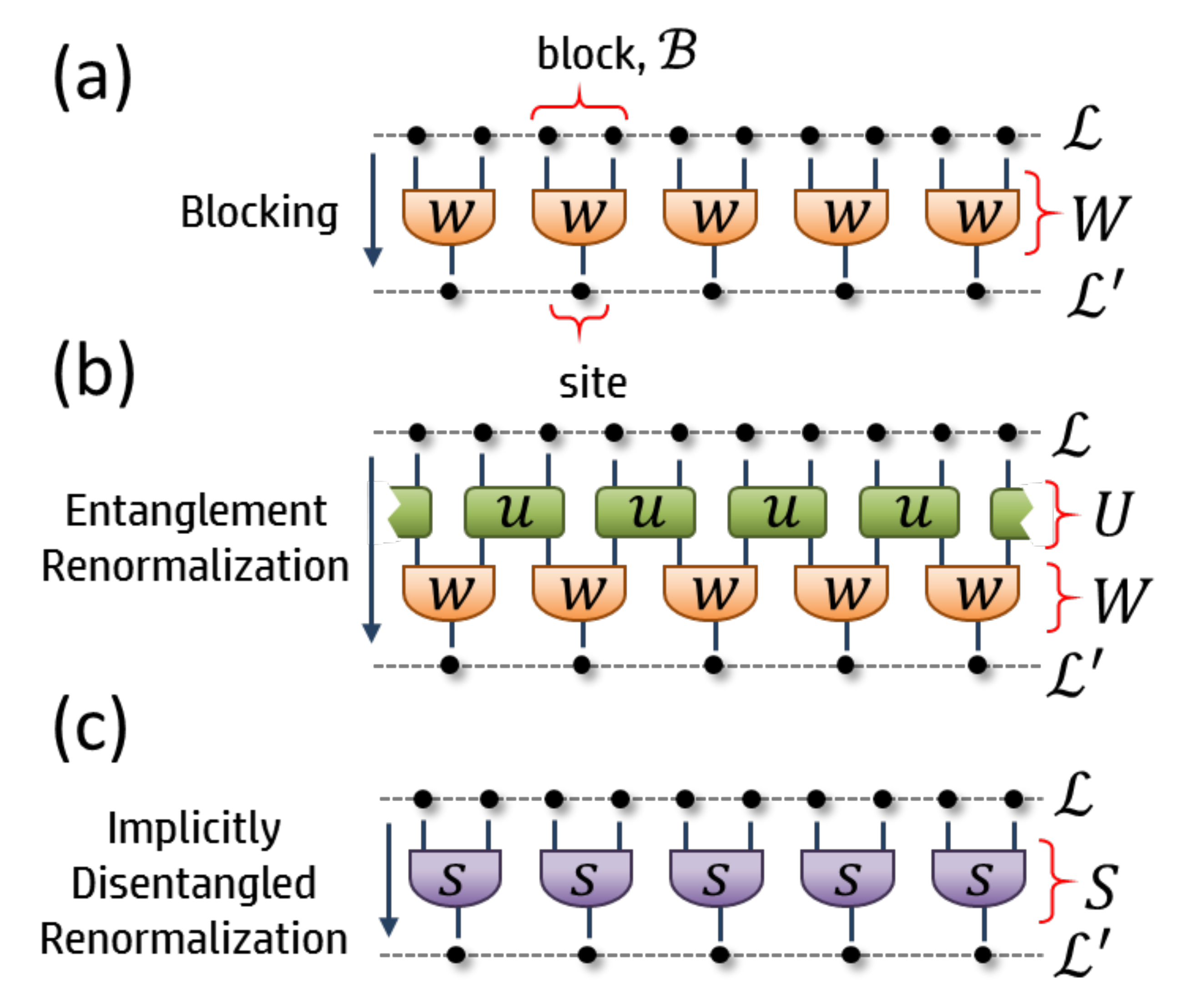}
\caption{(a) A coarse-graining transformation $W$ maps a quantum state on lattice $\L$ to a state on the coarser lattice $\L'$, where blocks $\B$ of 2 sites on $\L$ are each mapped to an effective site on $\L'$ via an isometry $w$. (b) Entanglement renormalization includes unitary disentanglers $u$ to remove entanglement between blocks as part of a coarse-graining transformation. (c) Implicitly disentangled renormalization (IDR) is implemented via (non-isometric tensors) $s$ that act locally within each block, yet also achieve the removal of entanglement between blocks.}
\label{fig:LatticeCG}
\end{center}
\end{figure}


In this manuscript we propose a coarse-graining transformation for quantum states that only uses operations local within each block, avoiding the use of disentanglers acting across block boundaries, yet that also retains the power of ER. The fundamental idea underlying this proposal is that the effect of certain unitary $u_{AB}$ operations, acting across two sub-regions $A$ and $B$ of a many-body state $\ket{\psi}$, can be replicated by local operators $s_A$, $s_B$ acting on each subregion separately, i.e. such that $u_{AB} \ket{\psi} = (s_A \otimes s_B) \ket{\psi}$. In the context that $u_{AB}$ is acting to disentangle the subregions, we refer to the use of local operators that reproduce the same effect as \emph{implicit disentangling}. This idea can be leveraged into an RG scheme for many-body quantum states, called implicitly disentangled renormalization (IDR), which we demonstrate to closely match the performance of ER.

The manuscript is organised as follows. First we review previous coarse-graining transformations, including \emph{blocking} and ER, before formulating implicit disentangling and testing its efficacy for the ground-state of a finite quantum system. We then apply IDR
 to coarse-grain the ground-state of an infinite critical system, represented by a matrix product state (MPS), which is shown to reach a scale-invariant fixed point.   

\textbf{Coarse graining transformations.---} We begin by recollecting previous coarse-graining transformations. Let $\mathcal{L}$ denote a $1D$ dimensional lattice made of $N$ sites, where each site is described by a Hilbert space $\mathbb{V}$ of finite dimension $d$, so that the vector space of the lattice is $\mathbb{V}_{ \mathcal{L}} \cong \mathbb{V}^{\otimes N}$. We shall consider transformations that map lattice $\mathcal L$ to a coarser lattice $\mathcal L'$ of $N/2$ sites, each with a vector space $\mathbb{V}'$ of dimension $\chi \le d^2$, so that $\mathbb{V}_{ \mathcal{L}'} \cong \mathbb{V}'^{\otimes N/2}$. The first transformation we discuss, $W:\mathbb V_{\mathcal{L}}  \mapsto \mathbb V_{\mathcal L'}$, is a \emph{blocking} transformation following the ideas of Wilson \cite{Wil3}, where blocks of two sites in $\L$ are each mapped to a site in lattice $\L'$ via an isometry $w$,
\begin{equation}
w:\mathbb V^{ \otimes 2}  \mapsto \mathbb V' ,\; \; \; \; w w^\dag = \mathbb I', \label{eq:M1}
\end{equation}
with $\mathbb I'$ the identity operator on $\mathbb{V}'$, such that $W = w^{\otimes N/2}$ as depicted in Fig. \ref{fig:LatticeCG}(a). The second transformation we consider is ER \cite{ER}, which augments the blocking transformation $W$ with a layer of unitary \emph{disentanglers} $U$, such that the full transformation is now $(WU):\mathbb V_{\mathcal{L}}  \mapsto \mathbb V_{\mathcal L'}$. Here $U = u^{\otimes N/2}$ is a product of two site unitary gates $u$,   
\begin{equation}
u:\mathbb V^{ \otimes 2}  \mapsto \mathbb V^{ \otimes 2}  ,\; \; \; u u^\dag = \mathbb I'^{ \otimes 2}, \label{eq:M2}
\end{equation}
which are enacted across the block boundaries as depicted in   Fig. \ref{fig:LatticeCG}(b).

Given a quantum state $\ket{\psi} \in \L$ the tensors comprising a coarse-graining transformation should be chosen such that the original state can be recovered as accurately as possible from the coarse-grained state (via a corresponding fine-graining transformation). For instance, in the case of the blocking transformation where the coarse-grained state on $\L'$ is defined $\ket{\psi'_\K} = W \ket{\psi}$, the recovered state is $\ket{\phi_\K} = W^\dag \ket{\psi'_\K} = W^\dag W \ket{\psi}$. Thus one should choose the isometries $w$ to maximise the scalar product,
\begin{equation}
\braket{\psi}{\phi_\K} = \bra{\psi} W^\dag W \ket{\psi}. \label{eq:M3}
\end{equation}
The identification of the optimal isometries, i.e. that maximise Eq. \ref{eq:M3}, was addressed by White during the formulation of the DMRG algorithm \cite{DMRG1, DMRG2}, who argued that an isometry $w$ should preserve the support of the local reduced density matrix $\rho_\B$ on the block $\B$ on which it acts. More precisely, if ${\rho _\B} = \sum\nolimits_k {\lambda _k}\ket{{v_k}}  \bra{{v_k}}$ with the eigenvalues ordered, $\lambda_{k} \ge \lambda_{k+1}$, then one should choose the isometry $w = \textrm{span}\{v_1, v_2, \ldots, v_\chi \}$. Similarly 
for ER, where the coarse-grained state on $\L'$ is given as $\ket{\psi'_\ER} = W U \ket{\psi}$, the isometries $w$ and disentanglers $u$ should be chosen to maximise the scalar product between the initial state $\ket{\psi}$ and the recovered state $\ket{\phi_\ER} =  U^\dag W^\dag \ket{\psi'_\ER} = U^\dag W^\dag W U \ket{\psi}$, i.e. such that,
\begin{equation}
\braket{\psi}{\phi_\ER} = \bra{\psi} U^\dag W^\dag W U \ket{\psi}, \label{eq:M4}
\end{equation}
is maximised. However, in general an exact solution for isometries $w$ and disentanglers $u$ maximising Eq. \ref{eq:M4} is not known, and one must rely on a variational approach \cite{ER,ERalg,ERferm,ERbose} to optimise the tensors. 

\textbf{Implicit disentangling.---}
We now introduce \emph{implicitly disentangled renormalization} (IDR) as a coarse-graining transformation, $S:\mathbb V_{\mathcal{L}}  \mapsto \mathbb V_{\mathcal L'}$, with the initial lattice $\L$ and coarser lattice $\L'$ as defined previously.
 The transformation $S$ is composed of a product of (non-isometric) tensors, $S = s^{\otimes N/2}$, such that each tensor $s$ maps a block of two sites in $\L$ to a single site in $\L'$, 
\begin{equation}
s:\mathbb V^{ \otimes 2}  \mapsto \mathbb V', \label{eq:M5}
\end{equation}
as depicted in Fig. \ref{fig:LatticeCG}(c). Thus far the transformation $S$ is equivalent to the blocking transformation $W$ depicted in Fig. \ref{fig:LatticeCG}(a), aside from lacking isometric constraints. The key difference, however, now comes with how the tensors in $S$ are chosen. As before, we desire that the approximation to an initial state $\ket{\psi}\in \L$ should be recoverable from the coarse-grained state $\ket{\psi'_\IDR} = S \ket{\psi}$, but now we allow the fine-graining transformation to be entirely distinct from $S$, rather than simply its conjugate. More precisely, we include unitary gates in the fine-graining transformation such that it mimics the structure of ER; the recovered state is thus defined as $\ket{\phi_\IDR} =  U^\dag W^\dag \ket{\psi'_\IDR} = U^\dag W^\dag S \ket{\psi}$. The tensors in $S$, $W$ and $U$ should be optimised to maximise the fidelity, 
\begin{equation}
F\left( {\psi ,{\phi _\IDR}} \right) = \frac{\braket{\phi _\IDR}{\psi} \braket{\psi}{\phi _\IDR}}{\braket{\phi _\IDR}{\phi _\IDR}}, \label{eq:M6}
\end{equation}
where the normalization ${\braket{\phi _\IDR}{\phi _\IDR}}$ has been included in the denominator since the tensors in $S$, which are unconstrained, can change the norm of $\ket{\phi _\IDR}$ arbitrarily. Notice that, if the fidelity in Eq. \ref{eq:M6} was unity, it would imply that the coarse-graining implemented by $S$ was equivalent to one implemented by ER, i.e. that $\ket{\psi'_\IDR} = S \ket{\psi} = W U \ket{\psi}$. Thus the key difference between a blocking transformation $W$ and one based upon implicit disentangling $S$ can be understood: each isometry in the blocking $W$ must retain all degrees of freedom that are entangled outside of the block, while with implicit disentangling the tensors in $S$ can truncate entangled degrees of freedom, and hence reduce the entanglement between blocks, provided that the entanglement can be restored by the fine-graining transformation. However it should be realised that IDR does not lead to a new variational ansatz for quantum states, as the fine-graining transformation of IDR is still equivalent to that of ER, and hence describes a multi-scale entanglement renormalization ansatz (MERA) \cite{MERA}, see Sect. C of the supplementary material for further details.

\begin{figure}[!t!b]
\begin{center}
\includegraphics[width=7.0cm]{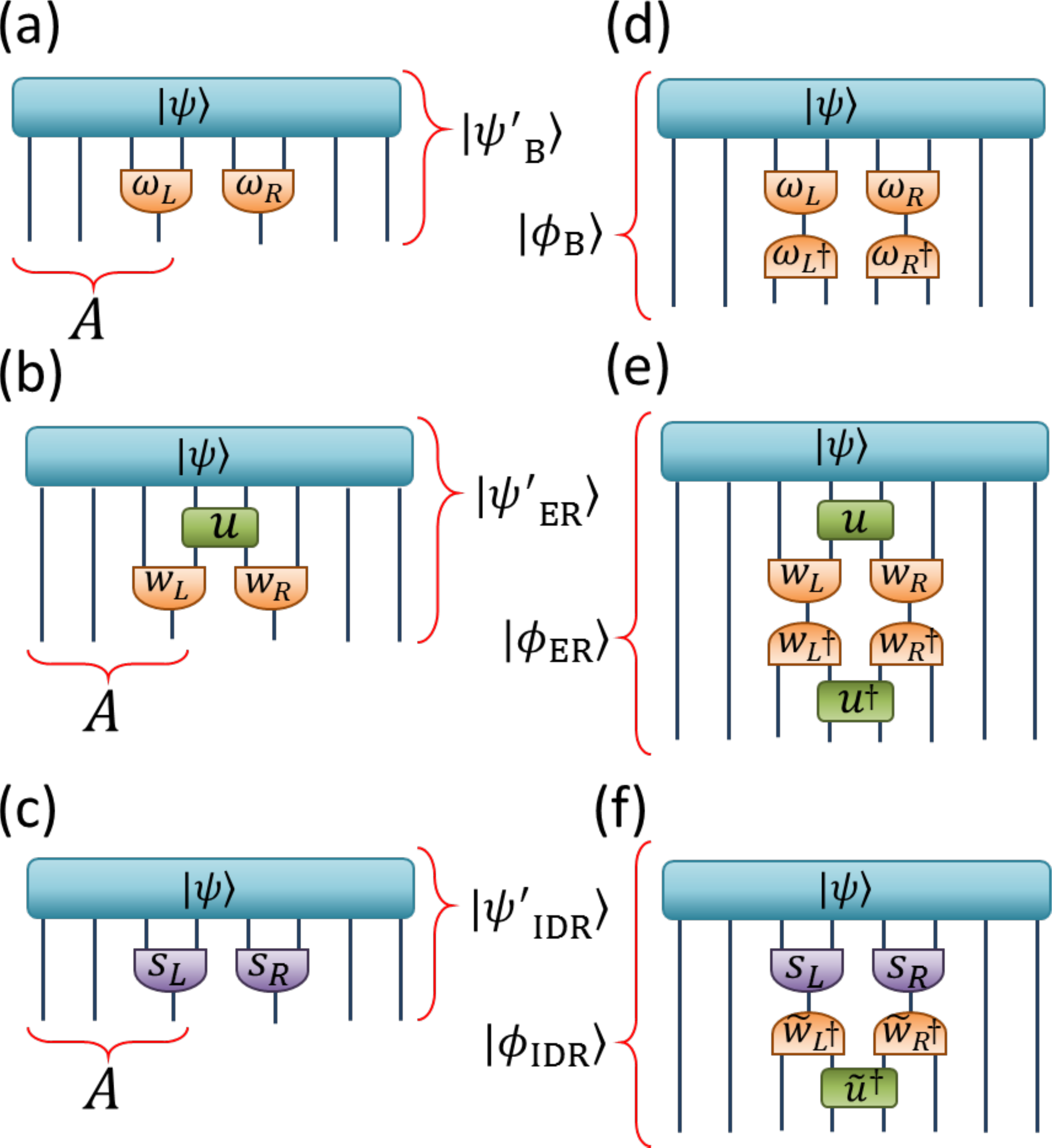}
\caption{(a-c) Coarse-grained states resulting from blocking, $\ket{\psi'_\K}$, entanglement renormalization $\ket{\psi'_\ER}$ and implicitly disentangled renormalization $\ket{\psi'_\IDR}$. (d-f) Approximations to the initial state $\ket{\psi}$ are recovered by fine-graining each of the states from (a-c).} 
\label{fig:Finite}
\end{center}
\end{figure}

\begin{table}[!htbp]
\begin{tabular}{||l|c||l|c||}
  \hline
                       & $S_A(\psi)$ &                     & $\varepsilon(\phi)$ \\ \hline
     $~$ $\ket{\psi}$ $~$   &  $~$ 1.179 $~$ & & \\
   $~$ $\ket{\psi'_\K}$  $~$ &  $~$ 1.175 $~$ & $~$ $\ket{\phi_\K} $~$ $   &  $~$ $3.12 \times 10^{-3}$ $~$ \\ 

   $~$ $\ket{\psi_\ER'}$ $~$ &  $~$ 0.976 $~$ & $~$ $\ket{\phi_\ER} $~$ $  &  $~$ $3.83 \times 10^{-6}$ $~$ \\

   $~$ $\ket{\psi'_\IDR}$ $~$ &  $~$ 0.976 $~$ & $~$ $\ket{\phi_\IDR} $~$ $  &  $~$ $4.05 \times 10^{-6}$ $~$ \\

  \hline
  \end{tabular} 
  \caption{The half-chain entanglement entropy $S_A$ and error in the fidelity, $\varepsilon(\phi) = 1 - F(\psi,\phi)$, with the initial state $\ket{\psi}$ for the states depicted in Fig. \ref{fig:Finite}.}
  \label{tab:Finite}
\end{table}

\textbf{Example: finite spin-chain.---} 
As a first test of implicit disentangling we study the ground state of the quantum critical Ising model, Hamiltonian $H = \sum\nolimits_r \left(Z_r - X_r X_{r+1} \right)$ with $Z$ and $X$ as Pauli matrices, on a periodic system of $N=24$ sites, which has been obtained through exact diagonalization. Before starting, we perform a preliminary blocking of every 3 spins into a single site of local dimension $d=8$, and then work with the ground state $\ket{\psi}$ on the (blocked) lattice $\L$ of 8 sites.

In order to compare the performance of the different RG schemes, we coarse-grain two neighbouring blocks $\B_L$ and $\B_R$ from $\L$, each of $2$ sites, into effective sites of dimension $\chi = 8$ (which have been truncated from the initial dimension of $d^2 = 64$). In the blocking scheme, two isometries are used to obtain the coarse-grained state, $\ket{\psi'_\K}  = \left( {{\omega_L} \otimes {\omega_R}} \right) \ket{\psi}$, see Fig. \ref{fig:Finite}(a), while with ER a 2-site disentangler $u$ is enacted between the blocks before the isometries, such that the coarse-grained state is $\ket{\psi'_\ER}  = \left( {{w_L} \otimes {w_R}} \right) u \ket{\psi}$, see Fig. \ref{fig:Finite}(b). With IDR the coarse-graining is implemented by tensors $s_L$ and $s_R$ which act locally within $\B_L$ and $\B_R$ respectively, $\ket{\psi'_\IDR}  = \left( {{s_L} \otimes {s_R}} \right) \ket{\psi}$, see Fig. \ref{fig:Finite}(c). The states recovered by fine-graining are $\ket{\phi_\K}  = ( {{\omega_L^\dag} \otimes {\omega_R^\dag}} ) \ket{\psi'_\K}$ for blocking, $\ket{\phi_\ER}  =u^\dag ( {{w_L^\dag} \otimes {w_R^\dag}} ) \ket{\psi'_\ER}$ for ER and $\ket{\phi_\IDR}  =\tilde u^\dag ( {{\tilde w_L^\dag} \otimes {\tilde w_R^\dag}} ) \ket{\psi'_\IDR}$ for IDR, see also Fig. \ref{fig:Finite}(d-f). In all three cases, the tensors comprising the transformations are optimised by maximising the fidelity of the recovered states with the initial state $\ket{\psi}$, see Sect. B of the supplementary material for additional details. The results, shown in Table. \ref{tab:Finite}, demonstrate the validity of implicit disentangling. Namely, we find that (i) the coarse-grained state resulting from implicit disentangling $\ket{\psi'_\IDR}$ exhibits the same reduction in entanglement entropy for the half-chain as that obtained with entanglement renormalization $\ket{\psi'_\ER}$, and (ii) that the accuracy of IDR, measured by the fidelity with which the initial state can be recovered from the coarser state, closely matches that of ER. 

\begin{figure}[!t!b]
\begin{center}
\includegraphics[width=8.5cm]{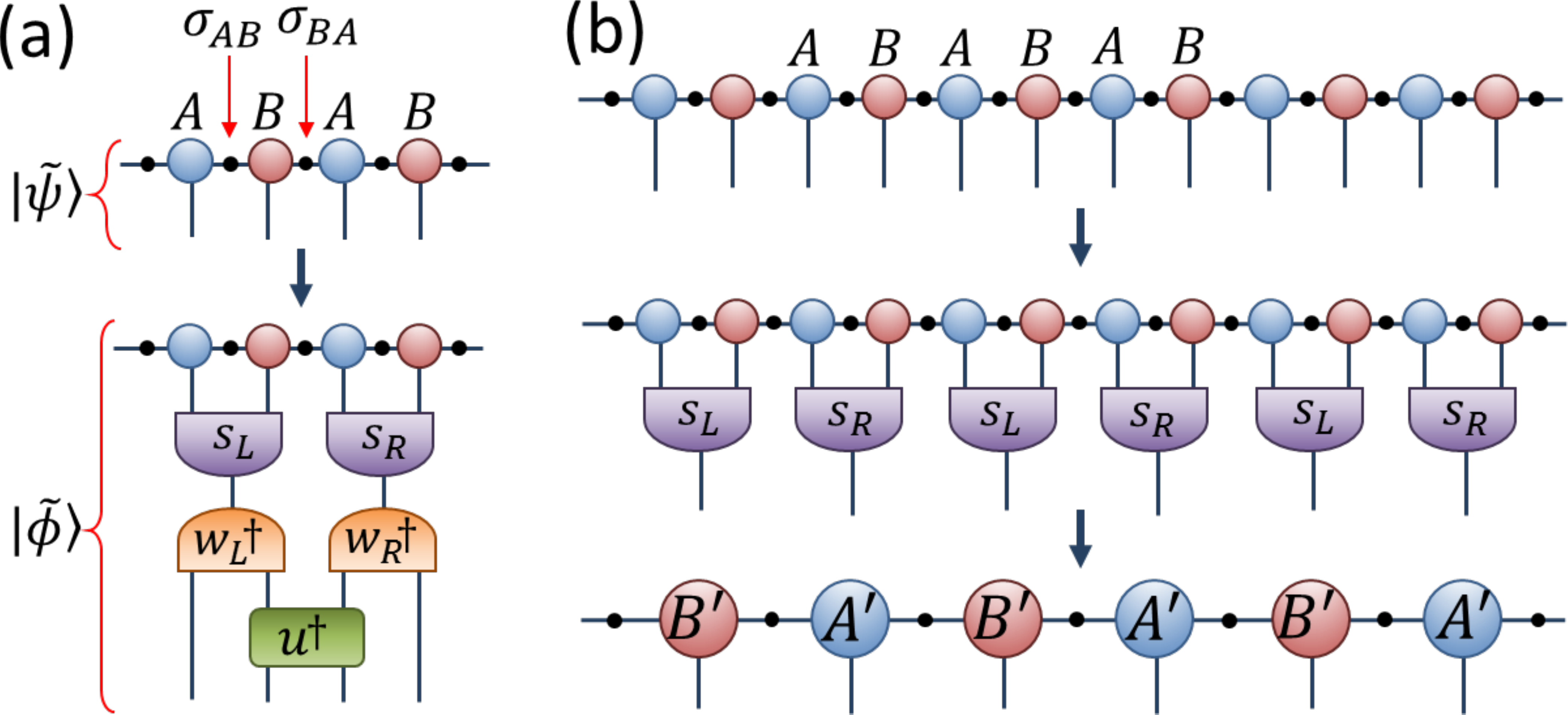}
\caption{(a) State $\ket{\tilde \psi}$ is defined from a four-site section of an MPS, while $\ket{\tilde \phi}$ is the state recovered after coarse-graining by tensors $s_L$ and $s_R$, then fine-graining by $w_L$, $w_R$ and $u$. (b) The MPS is coarse-grained via tensors $s_L$ and $s_R$, yielding a new MPS on a coarser lattice.  }
\label{fig:MPS}
\end{center}
\end{figure}

\textbf{Example: infinite spin-chain.---} 
As a second test of IDR we study an infinite system at criticality with the goal of demonstrating, as has been well established with ER \cite{ER,ERalg,ERferm,ERbose}, that IDR can achieve a sustainable RG flow (i.e. maintain a small truncation error over repeated RG steps) and can also capture scale-invariance \cite{Crit2, Crit3, Crit4, Crit5}. In particular we address the ground state of the quantum critical Ising model in the thermodynamic limit, which is obtained using iTEBD \cite{TEBD1,TEBD2} to optimise a matrix product state (MPS) \cite{MPS1,MPS2} of bond dimension $\chi = 100$ to within a relative energy error of $\delta E < 10^{-10}$. Before starting we perform a preliminary blocking of every 4 spins into a single site, obtaining an MPS \ket{\psi} of local dimension $d=16$, which is then brought into cannonical form. This MPS has a 2-site unit cell consisting of alternating tensors $A$ and $B$, interspersed with index weights $\sigma_{AB}$ and $\sigma_{BA}$, as depicted in Fig. \ref{fig:MPS}(a).  

The first step of the IDR iteration is to optimise the fidelity between the initial state $\ket{\psi}$ and the recovered state $\ket{\phi} = U^\dag W^\dag S \ket{\psi}$. Here we repeat the unit cell from the previous example, taking $S = (s_L \otimes s_R)^{\otimes \infty}$, $U = (\mathbb I \otimes u \otimes \mathbb I)^{\otimes \infty}$, and $W = (w_L \otimes w_R)^{\otimes \infty}$. Notice that this implementation uses half the number of unitaries in $U$ as originally depicted in Fig. \ref{fig:LatticeCG}(b), such that the fine-graining transformation relates to a modified binary MERA \cite{Crit4} as opposed to a standard binary MERA \cite{MERA}. This is done in order to simplify the optimisation; in particular the tensors $\{s_L, s_R, w_L, w_R, u \}$ can be optimised by maximizing the fidelity of a 4-site segment $\ket{\tilde \psi}$ of the MPS with the transformed state $\ket{\tilde \phi}$, see Fig. \ref{fig:MPS}(a) and Sect. B of the supplementary material for details. The MPS is coarse-grained using the optimised tensors $s_L$ and $s_R$, see Fig. \ref{fig:MPS}(b), to obtain a coarser (2-site unit cell) MPS. 
 The iteration of IDR is repeated many times, while maintaining a local dimension of $\chi=16$ at each RG step, thus generating a sequence of increasingly coarse-grained MPS. 

\begin{figure}[!t!b]
\begin{center}
\includegraphics[width=7.5cm]{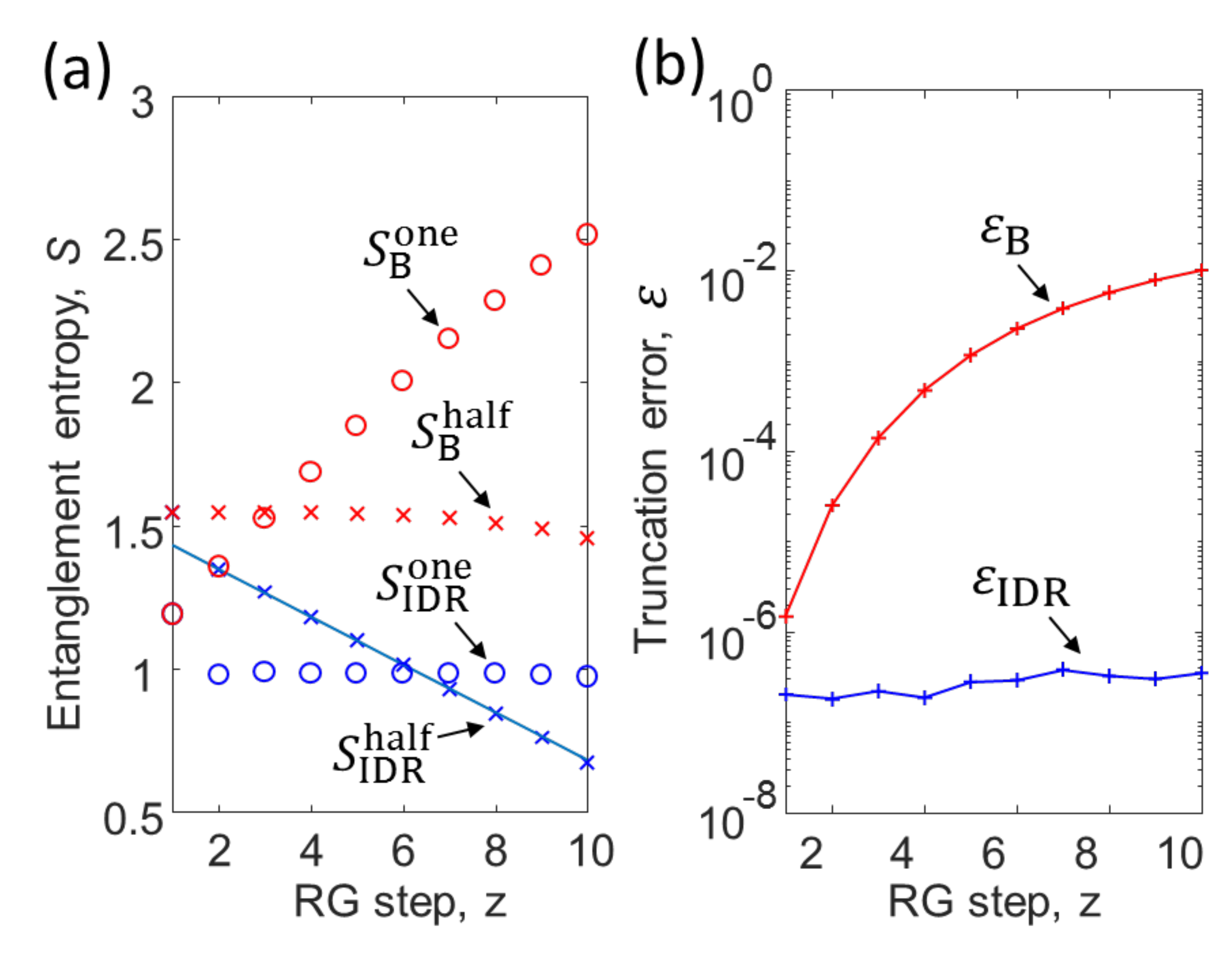}
\caption{Comparison of coarse-graining using blocking B versus IDR for (an MPS approximation to) the ground state of the quantum critical Ising model. (a) One-site entropies $S^\textrm{one}$ and half-chain entropies $S^\textrm{half}$. The solid line represents a fit of the decrease in entanglement predicted from theory \cite{Ent1,Ent2}, ${S_z} - {S_{z + 1}} = c/6$, with $c = 0.5$ as the central charge of the critical Ising model. (b) Local truncation errors, $\varepsilon = 1 - F$ with the fidelity $F$ as defined in Eq. \ref{eq:M6}, as a function of RG step (while keeping a fixed local dimension of $\chi=16$).}
\label{fig:Numeric}
\end{center}
\end{figure}

The results comparing the coarse-graining with IDR against a standard blocking transformation are shown in Fig. \ref{fig:Numeric}. As expected the blocking transformation breaks down after repeated RG steps, incurring successively larger truncation errors $\varepsilon_\K$, due to the growth in local entanglement $S_\K^\textrm{one}$ with each RG step. In contrast the transformation from IDR is sustainable over repeated iterations, without significant increase in truncation error $\varepsilon_\IDR$, due to the proper removal of short-ranged entanglement at each RG step as evidenced by a steady one-site entropy $S_\IDR^\textrm{one}$. Indeed we find that scale-invariance is fully realized with IDR; after $z=4$ initial RG steps the state becomes locally invariant \cite{LocalInv} with respect to the RG iteration, such that the tensors $\{s_L, s_R, w_L, w_R, u\}$ obtained in one RG iteration are identical to those obtained in the previous iteration up to very small errors \cite{FiniteCorr}. It follows that the isometries $w_L$, $w_R$ and unitaries $u$ optimised with IDR define a scale-invariant MERA \cite{Crit2, Crit3, Crit4, Crit5, ERalg} for the critical ground state. This MERA is found to have a relative energy error of $\delta E < 2.4\times 10^{-8}$, close to the energy error $\delta E < 1.1\times 10^{-8}$ of an equivalent $\chi=16$ MERA obtained through variational energy minimization \cite{Crit2}, further evidencing the validity of implicit disentangling. Finally, as demonstrated in Sect. C of the supplementary material, the conformal data extracted from this MERA is seen to accurately reproduce the Ising CFT \cite{CFT1,CFT2}, confirming that IDR is properly capturing the scale-invariant fixed point. 

\textbf{Discussion.---} The proposed method of implicit disentangling, which removes entanglement between blocks in a quantum state using local operators within each block, is seen to match the performance of unitary disentangling for quantum ground states. 
This is quite remarkable; while it is possible to construct examples where both methods of disentangling are exactly equivalent in their result, there are also counterexamples in which unitary disentangling is clearly more powerful (see Sect. A of the supplementary material). Indeed, the success of implicit disentangling may offer greater insight into the nature of entanglement in ground states in general.

Given that IDR inherits much of the simplicity of a blocking transformation, while still retaining the disentangling power of ER, it has many useful applications to tensor network algorithms. Here we outline just a few. Already demonstrated is a simple algorithm to translate an MPS into a MERA from which, if the MPS approximates the ground state of a critical system, the conformal data can be easily extracted. A distinct advantage of IDR over ER is that the form of the MPS is maintained under coarse-graining, as seen in Fig. \ref{fig:MPS}(b), whereas ER would require re-truncation of the MPS at every step. This algorithm could also be generalised to higher dimension, where it may be useful to contract PEPS \cite{PEPS1, PEPS2, PEPS3}. In particular, whereas previous methods for contracting a PEPS $\ket{\psi}$ involve contracting $\braket{\psi}{\psi}$ and thus squaring virtual dimension, the method with IDR would be applied to the state $\ket{\psi}$ directly, potentially leading to significant computational gains. 
In contrast, it is not even known how ER could be applied to coarse-grain a PEPS, due to the difficulty recovering the form of the PEPS after applying a unitary disentangler to a $2\times 2$ plaquette, as discussed further in Sect. D of the supplementary material. Another application of IDR would be towards tensor network renormalization (TNR) \cite{TNR1, TNR2, TNR3, TNR4} methods. As argued in Sect. D of the supplementary material, TNR algorithms can be simplified and their computational cost reduced by incorporating implicit disentangling. Moreover, the use of implicit disentangling could make TNR for $3D$ tensor networks, necessary for the study of $2D$ quantum systems, significantly more viable by allowing the structure of the network to be preserved under disentangling.

The author thanks David Poulin and Matthew Fishman for useful comments.

\newpage

\textbf{$~~~~~~~~~$SUPPLEMENTAL MATERIAL}

\section{Section A: Examples of implicit disentangling}
\begin{figure}[htb]
\begin{center}
\includegraphics[width=8.5cm]{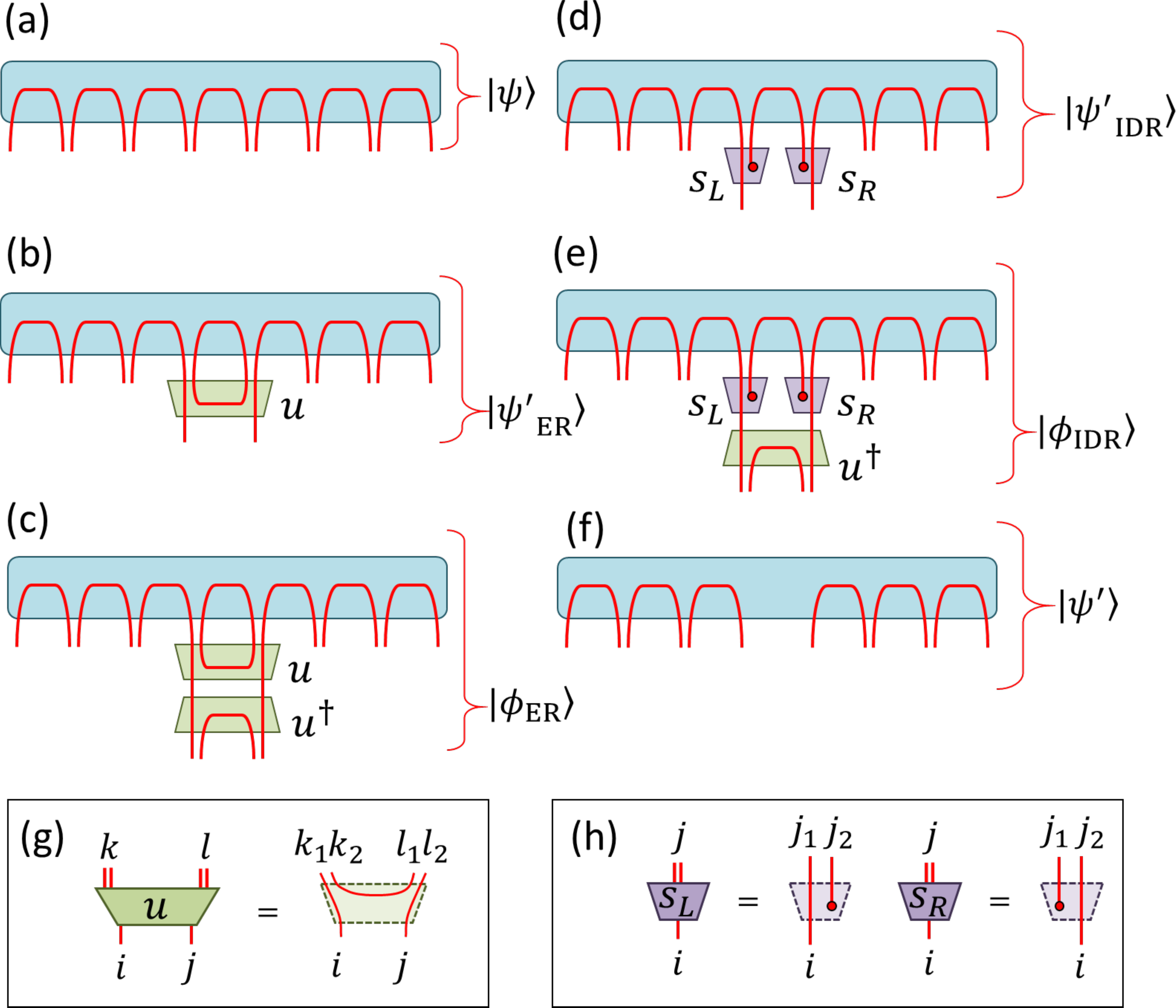}
\caption{(a) The quantum state $\ket{\psi}$ is a product of nearest neighbour singlets. (b) The isometry $u$ acts to disentangle $\ket{\psi}$. (c) The conjugate pair of isometries leaves state $\ket{\psi}$ invariant. (d) The state $\ket{\psi}$ is disentangled by tensors $s_L$ and $s_R$. (e) The initial state $\ket{\psi}$ is recovered through application of $u^\dag$. (f) The disentangled state $\ket{\psi'} = \ket{\psi'_\ER} = \ket{\psi'_\IDR}$. (g) Definition of the isometry $u$, see also Eq. \ref{eq:A1}. (h) Definition of tensors $s_L$ and $s_R$, see also Eq. \ref{eq:A2}. } 
\label{fig:ExactDisentangle}
\end{center}
\end{figure}

In this section we explore implicit disentangling in the context of toy models, providing one example where the effect of unitary disentangling can be exactly reproduced and one example where it can not. Let $\L$ be a $1D$ lattice where each site is described by a pair of qubits, and consider the quantum state $\ket{\psi} \in \L$ where one qubit from each site is in an entangled state, $\tfrac{1}{\sqrt{2}}(\ket{00} + \ket{11})$, with the site to the left and the other qubit is in an equivalent entangled state with the site to the right, see also Fig. \ref{fig:ExactDisentangle}(a). Using the ideas of entanglement renormalization, the state $\ket{\psi}$ can be fully disentangled (between a left/right partition) via an isometric disentangler $u_{ijkl}$ with indices $i,j\in\{1,2 \}$ and $k,l\in\{1,2,3,4 \}$, that is defined,
\begin{equation} 
{u_{ij({k_1}{k_2})({l_1}{l_2})}} = \tfrac{1}{{\sqrt 2 }}{\delta _{i{k_1}}}{\delta _{{k_2}{l_1}}}{\delta _{j{l_2}}}, \label{eq:A1}
\end{equation}
see also Fig. \ref{fig:ExactDisentangle}(g). Here we use a double index notation such that $k = (k_1 k_2)$ has the meaning $k = k_1 + 2(k_2 -1)$, where the single index runs over values $k_1,k_2 \in \{1,2\}$. The coarse-grained state $\ket{\psi'_\ER} =  u \ket{\psi}$ is now disentangled between the sites which $u$ was acted upon, see Fig. \ref{fig:ExactDisentangle}(b), whilst the fine-graining $\ket{\phi_\ER} = u^\dag \ket{\psi'_\ER} = u^\dag u \ket{\psi}$ still exactly recovers the initial state $\ket{\psi}$, see Fig. \ref{fig:ExactDisentangle}(c). We now demonstrate that the same disentangling of $\ket{\psi}$ can be achieved via implicit disentangling. Tensors $(s_L)_{ij}$ and $(s_R)_{ij}$, with indices $i\in\{1,2 \}$ and $j\in\{1,2,3,4 \}$, are defined,
\begin{align} 
{\left( s_L \right)_{i({j_1}{j_2})}} &= 2^{(1/4)} {\delta _{i{j_1}}}{\delta _{1{j_2}}},\nonumber \\
{\left( s_R \right)_{i({j_1}{j_2})}} &= 2^{(1/4)} {\delta _{1{j_1}}}{\delta _{i{j_2}}}, \label{eq:A2}
\end{align}
where the double index notation has been used, see also Fig. \ref{fig:ExactDisentangle}(h). The coarse-grained state $\ket{\psi'_\IDR} =  (s_L \otimes s_R) \ket{\psi}$ is now seen to exactly match that obtained through unitary disentangling $\ket{\psi'_\ER}$, as depicted in Fig. \ref{fig:ExactDisentangle}(e), although the mechanism with which the entanglement was removed is very different. With implicit disentangling, the qubits that were entangled across the partition are individually truncated from the state via the tensors $(s_L)_{ij}$ and $(s_R)_{ij}$. However the fine-graining $\ket{\phi_\IDR} = u^\dag \ket{\psi'_\IDR} = u^\dag (s_L \otimes s_R) \ket{\psi}$ again exactly recovers the initial state $\ket{\psi}$, see Fig. \ref{fig:ExactDisentangle}(e). 

\begin{figure}[!t!b]
\begin{center}
\includegraphics[width=8.6cm]{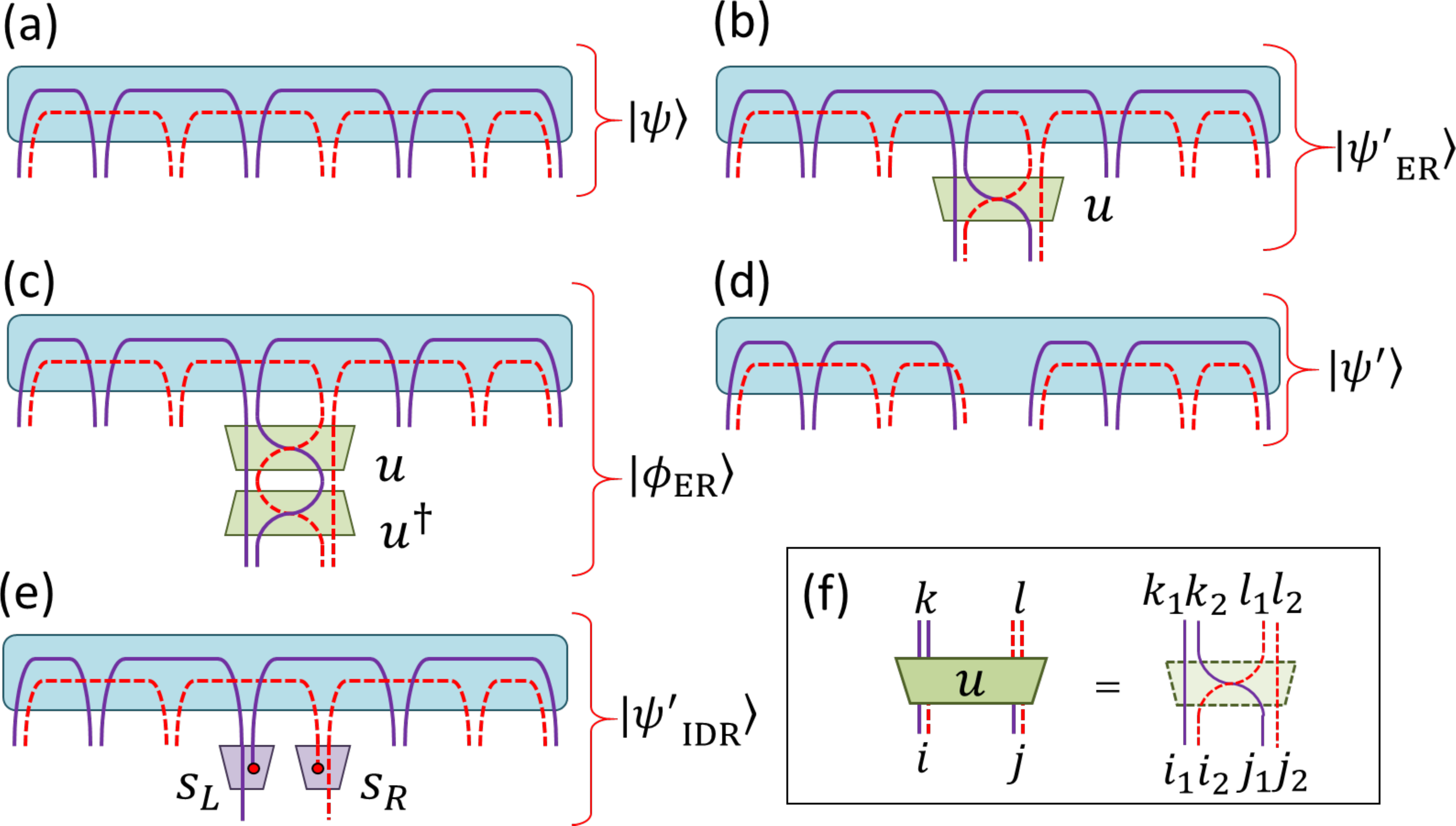}
\caption{(a) The quantum state $\ket{\psi}$ is a product of next-nearest neighbour singlets. (b) The unitary $u$ acts to disentangle $\ket{\psi}$. (c) The state is re-entangled via the action of $u^\dag$. (d) The disentangled state $\ket{\psi'} = \ket{\psi'_\ER}$. (e) Tensors $s_L$ and $s_R$ chosen to remove entanglement across the left/right partition. (f) Definition of the unitary $u$, see also Eq. \ref{eq:A3}. }
\label{fig:NoDisentangle}
\end{center}
\end{figure}

We now provide an example where the effect of unitary disentangling cannot be reproduced using implicit disentangling. Let $\L$ be a $1D$ lattice where each site is described by a pair of qubits, and consider the quantum state $\ket{\psi} \in \L$ where one qubit from each site is in an entangled state, $\tfrac{1}{\sqrt{2}}(\ket{00} + \ket{11})$, with next-nearest site to the left and one with the next-nearest site to the right, see also Fig. \ref{fig:NoDisentangle}(a). The state $\ket{\psi}$ can be fully disentangled (between a left/right partition) via a unitary disentangler $u_{ijkl}$ with indices $i,j,k,l\in\{1,2,3,4 \}$ that is defined,
\begin{equation} 
{u_{({i_1}{i_2})({j_1}{j_2})({k_1}{k_2})({l_1}{l_2})}} = {\delta _{i_1 {k_1}}}{\delta _{{i_2}{l_1}}}{\delta _{j_1{k_2}}}{\delta _{j_2{l_2}}}, \label{eq:A3}
\end{equation}
see also Fig. \ref{fig:NoDisentangle}(f), where we again use a double index notation such that $i = (i_1 i_2)$ has the meaning $i = i_1 + 2(i_2 -1)$. Notice that the coarse-grained state $\ket{\psi'_\ER} =  u \ket{\psi}$ is now disentangled across a left-right partition, see Fig. \ref{fig:NoDisentangle}(b), whilst the fine-graining $\ket{\phi_\ER} = u^\dag \ket{\psi'_\ER} = u^\dag u \ket{\psi}$ still exactly recovers the initial state $\ket{\psi}$, see Fig. \ref{fig:NoDisentangle}(c). However, in this case it is clear that this disentangling cannot be reproduced implicitly by a product of local operators, $(s_L \otimes s_R)$, that act on the same region as $u$. More precisely, although we can define the tensors,
\begin{align} 
{\left( s_L \right)_{i({j_1}{j_2})}} &= 2^{(1/4)} {\delta _{i{j_1}}}{\delta _{1{j_2}}},\nonumber \\ 
{\left( s_R \right)_{i({j_1}{j_2})}} &= 2^{(1/4)} {\delta _{1{j_1}}}{\delta _{i{j_2}}}, \label{eq:A4}
\end{align}
such that the state $\ket{\psi'_\IDR} = (s_L \otimes s_R)\ket{\psi}$ is fully disentangled across the partition, see also Fig. \ref{fig:NoDisentangle}(e), there does not exist an isometry $u$ (acting on the output sites of $s_L$ and $s_R$) that can restore the initial state as $u^\dag \ket{\psi'_\IDR}$. However, an isometry $u$ with support on four sites, including the output sites of $s_L$ and $s_R$ and their neighbouring sites, could restore the initial state $\ket{\psi}$.

\begin{figure}[!t!b]
\begin{center}
\includegraphics[width=8.6cm]{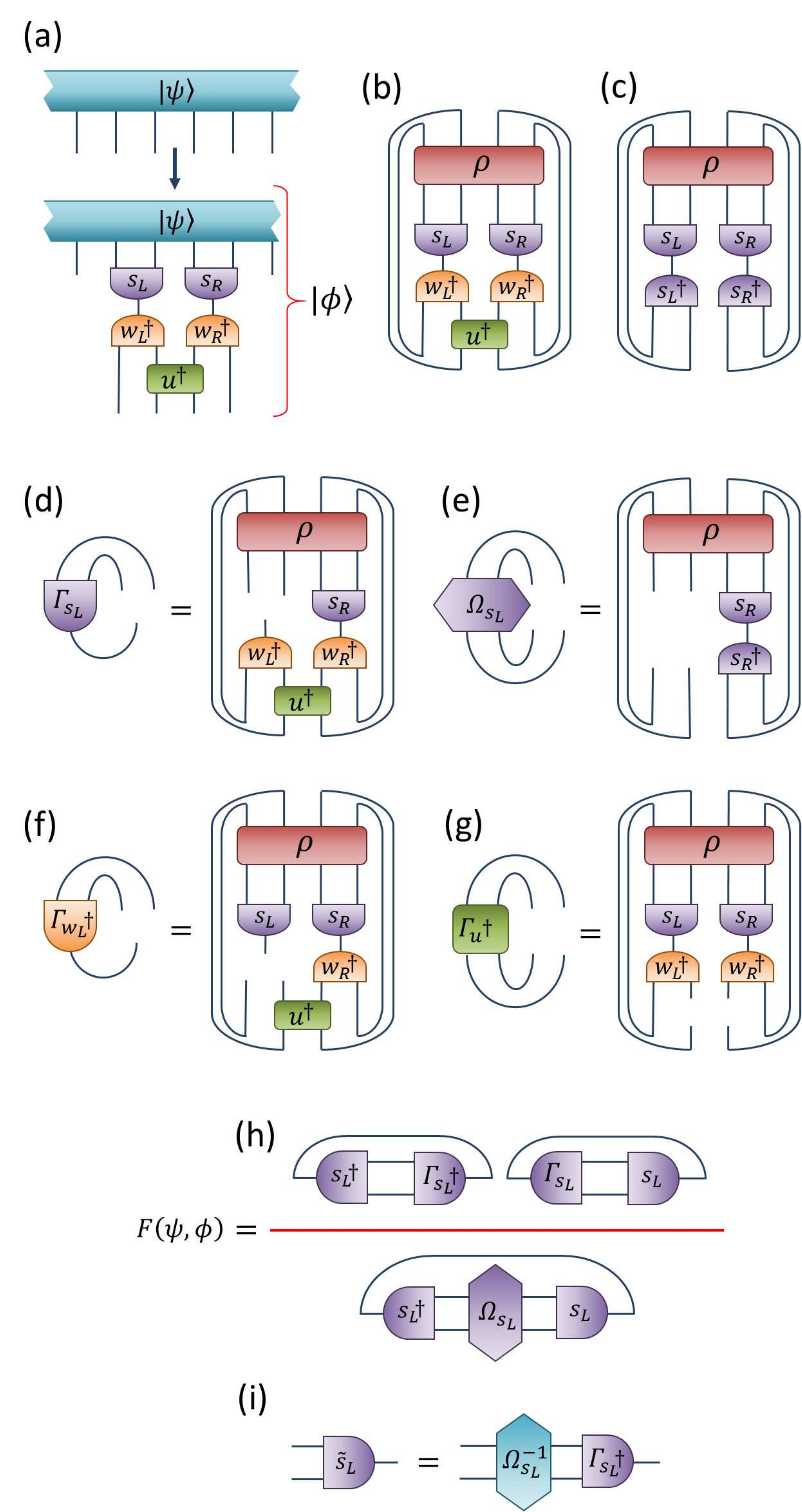}
\caption{(a) The initial state $\ket{\psi}$ and the state $\ket{\phi}$ recovered after coarse-graining with $s_L$, $s_R$ and then fine-graining with $w_L$, $w_R$ and $u$. (b) Depiction of $\braket{\phi}{\psi}$, with $\rho$ the reduced density matrix from $\ket{\psi}$. (c) Depiction of $\braket{\phi}{\phi}$. (d) Tensor environment $\Gamma_{s_L}$ of $s_L$ from $\braket{\phi}{\psi}$. (e) Tensor environment $\Omega_{s_L}$ of $(s_L,s_L^\dag)$ from $\braket{\phi}{\phi}$. (f) Tensor environment $\Gamma {w_L^\dag}$ of $w_L^\dag$ from $\braket{\phi}{\psi}$. (g) Tensor environment $\Gamma_{u^\dag}$ of $u^\dag$ from $\braket{\phi}{\psi}$. (h) The fidelity $F\left( {\psi ,{\phi}} \right)$, see Eq. \ref{eq:B5}, expressed in terms of $s_L$ and its environments. (i) The fidelity is maximised by updating the tensor $s_L \rightarrow \tilde s_L$.}
\label{fig:Optimization}
\end{center}
\end{figure}

\section{Section B: Optimisation for implicit disentangling}
In this section we provide additional details for how the tensors defining an iteration of IDR can be optimised. Let $\mathcal{L}$ denote a $1D$ dimensional lattice where each site is described by a Hilbert space $\mathbb{V}$ of finite dimension $d$. We consider two tensors $s_L$ and $s_R$ that each map a pair of sites in $\L$ to an effective site of dimension $\chi \le d^2$,
\begin{equation}
s_L:\mathbb V^{ \otimes 2}  \mapsto \mathbb V',\;\;\;\; s_R:\mathbb V^{ \otimes 2}  \mapsto \mathbb V', \label{eq:B1}
\end{equation}
with $\mathbb V'$ the Hilbert space of an effective site. Similarly we consider a pair of isometries $w_L$ and $w_R$ that also map a pair of sites in $\L$ to an effective site
\begin{equation}
w_L:\mathbb V^{ \otimes 2}  \mapsto \mathbb V',\;\;\;\; w_R:\mathbb V^{ \otimes 2}  \mapsto \mathbb V', \label{eq:B2}
\end{equation}
which are constrained such that $w_L w_L^\dag = \mathbb{I}'$ and $w_R w_R^\dag = \mathbb{I}'$, with $\mathbb{I}'$ the identity operator on $\mathbb V'$. Finally we consider a unitary $u$ acting on a pair of sites in $\L$,
\begin{equation}
u:\mathbb V^{ \otimes 2}  \mapsto V^{ \otimes 2}, \label{eq:B3}
\end{equation}
which is constrained such that $u u^\dag = \mathbb{I}\otimes \mathbb{I}$. Let us define the recovered state,
\begin{equation}
\ket{\phi} = \big( \mathbb{I}\otimes u^\dag \otimes \mathbb{I} \big) \big( w_L^\dag \otimes w_R^\dag \big) \big( s_L \otimes s_R \big) \ket{\psi}, \label{eq:B4}
\end{equation}
see also Fig. \ref{fig:Optimization}(a). Our goal is to optimise the tensors $\{s_L, s_R, w_L, w_R, u \}$ in order to maximise the fidelity between the initial and recovered state,
\begin{equation}
F\left( {\psi ,{\phi}} \right) = \frac{\braket{\phi}{\psi} \braket{\psi}{\phi}}{\braket{\phi}{\phi }}, \label{eq:B5}
\end{equation}
where the normalization term ${\braket{\phi}{\phi }}$ in the denominator is necessary as the tensors $s_L$ and $s_R$ are unconstrained (such that they can change the norm of $\ket{\phi}$ arbitrarily). Notice that both $\braket{\phi}{\psi}$ and $\braket{\phi}{\phi}$ can be expressed as a finite tensor network in terms of the four-site reduced density matrix $\rho$ on which they act, see Fig. \ref{fig:Optimization}(b) and Fig. \ref{fig:Optimization}(c) respectively. In order to optimise the fidelity of Eq. \ref{eq:B5} we use an iterative update strategy, where single tensors are updated while the rest are held fixed. We begin by discussing how the tensors $s_L$ and $s_R$ associated to the implicit disentangling are updated. Let us define $\Gamma s_L$ as the environment of $s_L$ from $\braket{\psi}{\phi}$, see Fig. \ref{fig:Optimization}(d), and similarly define $\Omega s_L$ as the environment generated by removing both $(s_L, s_L^\dag)$ from $\braket{\phi}{\phi}$, see Fig. \ref{fig:Optimization}(e). Then the fidelity may be expressed as,
\begin{equation}
F\left( {\psi ,\phi } \right) = \frac{{\textrm{tr}\left( {s_L^\dag \Gamma _{{s_L}}^\dag } \right) \textrm{tr}\Big( {{\Gamma _{{s_L}}}{s_L}} \Big)}}{{\textrm{tr}\left( {s_L^\dag {\Omega _{{s_L}}}{s_L}} \right)}}, \label{eq:B6}
\end{equation}
see also Fig. \ref{fig:Optimization}(h), which we recognize as a generalized eigenvalue problem for $s_L$, i.e. a maximization for vector $x$ of the form $(x A x^\dag) / (x B x^\dag)$. It can then be shown that the dominant (generalised) eigenvector $\tilde s_L$ of Eq. \ref{eq:B6} is given as,
\begin{equation}
\tilde s_L = \left(\Omega _{{s_L}} \right)^{-1} \Gamma_{s_L}^\dag ,\label{eq:B7} 
\end{equation}
with $\Omega^{-1}$ as the matrix inverse of $\Omega$, see also Fig. \ref{fig:Optimization}(i). Note that, in practice, one should use the pseudoinverse of $\Omega$ to avoid the problem of zero eigenvalues. The same optimization strategy is also used for the update of $s_R$.

We now discuss how the isometries $w_L$, $w_R$ and unitary $u$ can be optimised. Firstly, since the normalization $\braket{\phi}{\phi}$ of state $\ket{\phi}$ is independent of these tensors, as seen in Fig. \ref{fig:Optimization}(c), it follows that the fidelity in Eq. \ref{eq:B6} is maximised by simply maximizing the scalar product $\braket{\psi}{\phi}$. Here one may use the standard optimization method of ER for isometric/unitary tensors \cite{ERalg}, based on the singular value decomposition (SVD) of their environment. For instance, in order to update the unitary $u$ one would first compute the environment $\Gamma_u$ from $\braket{\psi}{\phi}$ as depicted in Fig. \ref{fig:Optimization}(g), and then take the SVD of the environment,
\begin{equation}
\Gamma_u = V_1 S V_2^\dag, \label{eq:B8}
\end{equation}
with $V_1$ and $V_2$ as unitary matrices and $S$ as the diagonal matrix of singular values. The updated tensor $\tilde u$ is then chosen as,
\begin{equation}
\tilde u = V_2 V_1^\dag . \label{eq:B9}
\end{equation}

\begin{figure}[!t!b]
\begin{center}
\includegraphics[width=8.6cm]{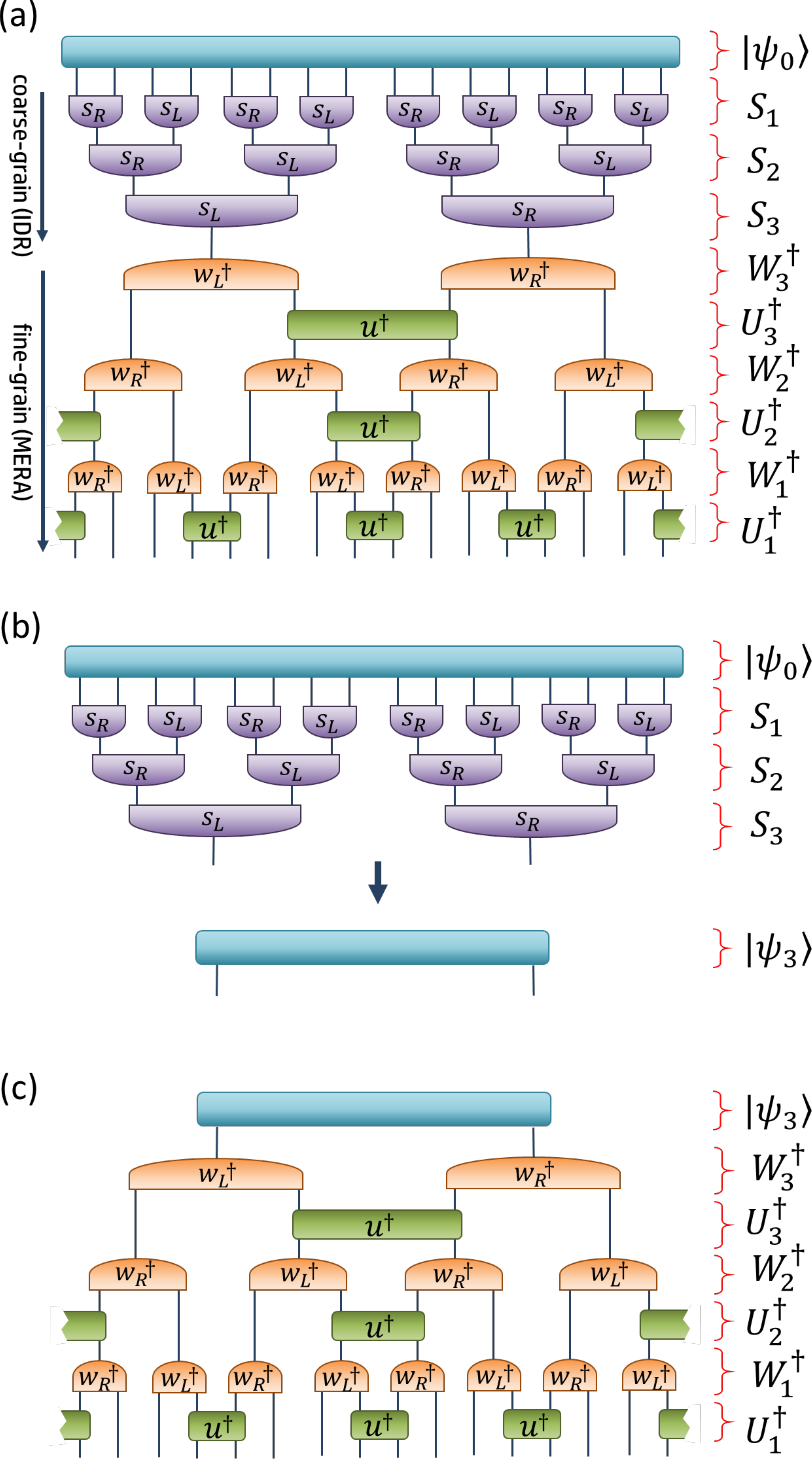}
\caption{(a) Given the quantum state $\ket{\psi_0}$ on a lattice of $N=16$ sites, the sequence of three coarse-graining transformations and the corresponding fine-graining transformations (assuming periodic boundaries) are optimised to leave the state approximately invariant, i.e. such that $\ket{\psi_0} \approx \big( U_1^\dag W_1^\dag \big) \big( U_2^\dag W_2^\dag \big) \big( U_3^\dag W_3^\dag \big) S_3 S_2 S_1 \ket{\psi_0}$. (b) The coarse-grained state $\ket{\psi_3}$ is obtained from three iterations of IDR applied to the initial state, $\ket{\psi_3} = S_3 S_2 S_1 \ket{\psi_0}$. (c) The coarse-grained state $\ket{\psi_3}$ in conjunction with the fine-graining transformation define a MERA approximation to the ground-state, $\ket{\psi_0} \approx \big( U_1^\dag W_1^\dag \big) \big( U_2^\dag W_2^\dag \big) \big( U_3^\dag W_3^\dag \big) \ket{\psi_3}$. }
\label{fig:Double}
\end{center}
\end{figure}

\section{Section C: MERA from implicit disentangling}
In this section we further detail how a MERA approximation to an initial state $\ket{\psi_0}$ is generated using IDR. Furthermore we demonstrate that the MERA obtained from an MPS approximation to the ground state of the quantum critical Ising model, as discussed in the main text, accurately reproduces the critical data characterizing the Ising CFT.

Each coarse-graining step of IDR involves optimising the fidelity between an initial state $\ket{\psi}$ and a recovered state $\ket{\phi} = U^\dag W^\dag S\ket{\psi}$, where $S$ and $W$ are both a product of tensors, that each map a block of sites to an effective site of a coarser lattice, and $U$ is a product of unitaries that are enacted across block boundaries. There are many different patterns with which one can organise the blocks and unitaries, which result in different forms of MERA. Here, as in the main text, we focus on the pattern formed from the unit cell depicted in Fig. \ref{fig:Optimization}(a) which gives a modified binary MERA \cite{Crit2}. Let us consider a quantum state $\ket{\psi_0}$ on a lattice of $N=16$ sites, and imagine a sequence of coarser states has been generated from $z=3$ iterations of IDR, 
\begin{equation}  
\left| {{\psi _0}} \right\rangle \mathop  \to \limits^{{S_1}} \left| {{\psi _1}} \right\rangle \mathop  \to \limits^{{S_2}} \left| {{\psi _2}} \right\rangle \mathop  \to \limits^{{S_3}} \left| {{\psi _3}} \right\rangle, 
\end{equation}
where $\ket{\psi_z} = S_{z} \ket{\psi_{z-1}}$ is a state on a lattice $\L_z$ of $N_z = 2^{4-z}$ sites. As previously discussed, each iteration is optimised such that such that 
\begin{equation}
U_z^\dag W_z^\dag \ket{\psi_z} \approx \ket{\psi_{z-1}},
\end{equation}
which further implies that,
\begin{equation}
\ket{\psi_0} \approx \big( U_1^\dag W_1^\dag \big) \big( U_2^\dag W_2^\dag \big) \big( U_3^\dag W_3^\dag \big) S_3 S_2 S_1 \ket{\psi_0},
\end{equation}
see also Fig. \ref{fig:Double}(a). It follows that the coarse-grained state $\ket{\psi_3}$, which is defined on a lattice $\L_3$ of $N_3 = 2$ sites, in conjunction with the unitary $U_z$ and isometric $W_z$ layers constitute a MERA approximation to the ground-state,
\begin{equation}
\ket{\psi_\MERA} = \big( U_1^\dag W_1^\dag \big) \big( U_2^\dag W_2^\dag \big) \big( U_3^\dag W_3^\dag \big) \ket{\psi_3},
\end{equation}
see Fig. \ref{fig:Double}(b-c). 

\begin{figure}[!t!b]
\begin{center}
\includegraphics[width=8.6cm]{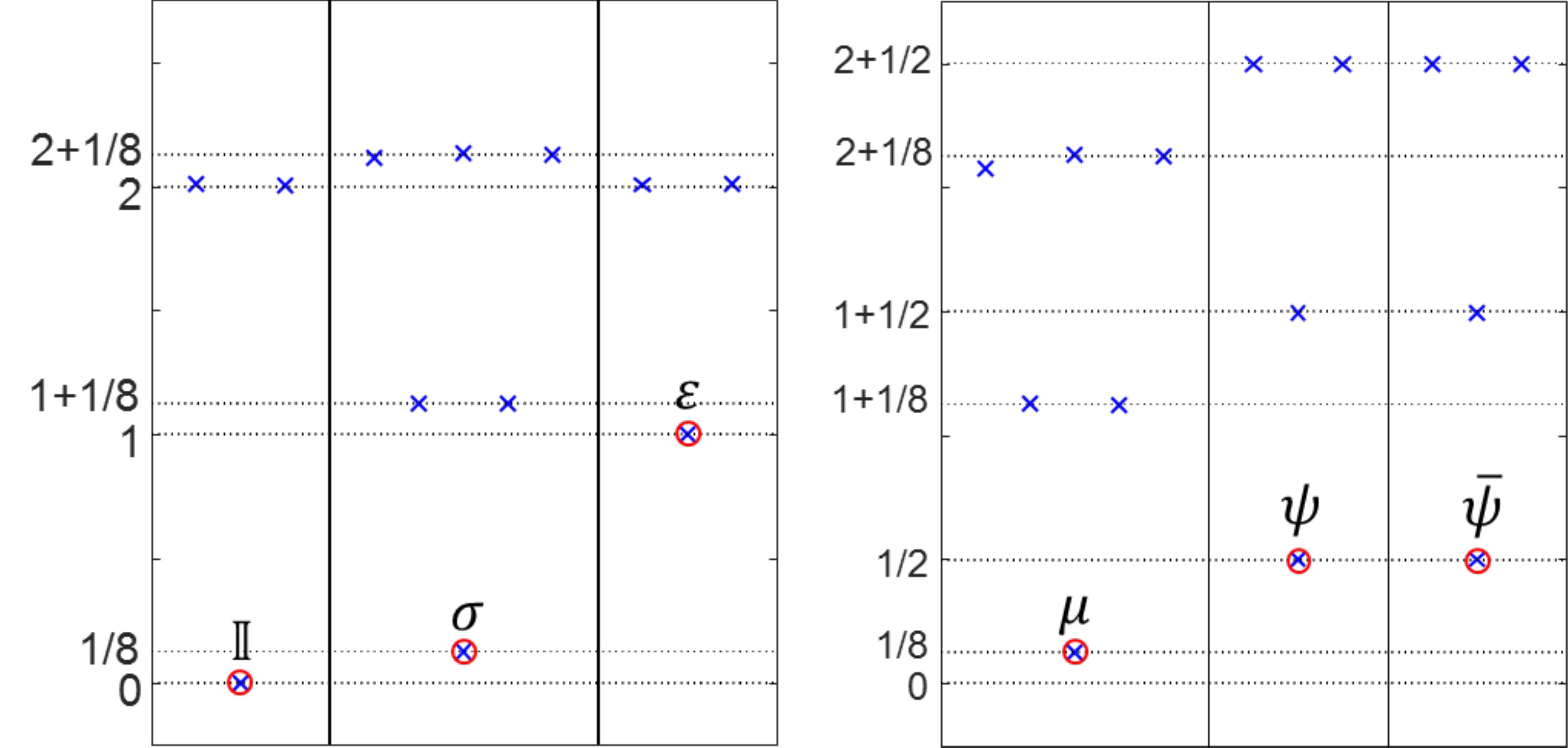}
\caption{Scaling dimensions of local and non-local scaling operators of the Ising CFT, computed from a $\chi=16$ scale-invariant MERA obtained with IDR. }
\label{fig:Scaling}
\end{center}
\end{figure}

In the main text, an MPS approximation to the ground-state of the critical Ising model on an infinite lattice was coarse-grained using IDR of bond dimension $\chi=16$. After $z=4$ initial coarse-graining iterations a scale-invariant fixed point was reached, such that the subsequent layers of the transformation were approximately equal to the previous ones, i.e such that $S_{z+1} \approx S_z$ for $z>4$. Thus a (modified binary) scale-invariant MERA approximation to the ground-state of the infinite critical Ising model is obtained, with $\{ U_1, W_1, U_2, W_2, U_3, W_3 \}$ as the transitional layers, and $\{ U_4, W_4 \}$ as the scale-invariant layer. The relative error in the energy of this MERA was evaluated at $\delta E = 2.4 \times 10^{-8}$, which is very close to the relative error $\delta E = 1.1 \times 10^{-8}$ of a $\chi=16$ MERA obtained through variational energy minimization \cite{Crit2}. One can extract the conformal data of the Ising CFT from this MERA using standard techniques \cite{Crit2,Crit3,Crit4}, involving diagonalizing the scaling superoperators associated to the scale-invariant layer. The results, again found to be of comparable accuracy to a $\chi=16$ MERA obtained through variational energy minimization, are shown in Fig. \ref{fig:Scaling} and Table \ref{tab:IsingExp} for the scaling dimensions, while Table \ref{tab:IsingOPE} shows the fusion coefficients. 

\begin{table}[!htbp]
\centering
\renewcommand{\arraystretch}{1.2}
\setlength\tabcolsep{5pt}
\begin{tabular}{ |l | l | l |}
\hline
$\Delta^{\mbox{\tiny exact}}$  & $~\Delta^{\mbox{\tiny MERA}}_{\mbox{\tiny $\chi=16$}}~$ & Error  \\
\hline
$\Delta_\sigma$=0.125       &  $~$ 0.125109   & 0.08 $\%$ \\
  $\Delta_\epsilon$=1       &  $~$ 1.000639   & 0.06 $\%$ \\
  $\Delta_\mu$=0.125        &  $~$ 0.125019   & 0.01 $\%$ \\
  $\Delta_\psi$=0.5         &  $~$ 0.500273   & 0.05 $\%$ \\
  $\Delta_{\bar\psi}$=0.5   &  $~$ 0.500273   & 0.05 $\%$ \\
  \hline\end{tabular}
  \caption[]{Scaling dimensions of the primary fields of the Ising CFT, computed from a $\chi=16$ scale-invariant MERA obtained with IDR.}
\label{tab:IsingExp}       
\end{table}

\begin{table}[!tbp]
\centering
\renewcommand{\arraystretch}{1.2}
\setlength\tabcolsep{5pt}
\begin{tabular}{ |l | l | l |}
\hline
$~~~~~C^{\mbox{\tiny exact}}$  & $~C^{\mbox{\tiny MERA}}_{\mbox{\tiny $\chi=16$}}~$ & error  \\
\hline
$C_{\epsilon, \sigma, \sigma} = 1/2$ & $~$  0.5021 & 0.42$\%$ \\
$C_{\epsilon, \mu, \mu} = -1/2$      & $~$ -0.4983 & 0.34$\%$    \\
$C_{\psi,     \mu,    \sigma} = \frac{e^{-i\pi/4}}{\sqrt{2}}$     &  $\frac{1.0132e^{-i\pi/4}}{\sqrt{2}}$ $~$ & 1.32$\%$ \\ 
   $C_{\bar\psi,     \mu,    \sigma} = \frac{e^{i\pi/4}}{\sqrt{2}}$  &  $\frac{1.0132e^{i\pi/4}}{\sqrt{2}}$ $~$  & 1.32$\%$   \\

   $C_{\epsilon, \psi, \bar\psi} = i$     &  $1.0083 i$  $~$ & 0.83$\%$ \\
   $C_{\epsilon, \bar\psi, \psi} = -i$    &  $-1.0083 i$ $~$ & 0.83$\%$   \\ \hline\end{tabular}
   \caption[]{Operator product expansion (OPE) coefficients for the local and non-local primary fields of the Ising CFT, computed from a $\chi=16$ scale-invariant MERA obtained with IDR.}
\label{tab:IsingOPE}       
\end{table}

\begin{figure}[!t!b]
\begin{center}
\includegraphics[width=8.6cm]{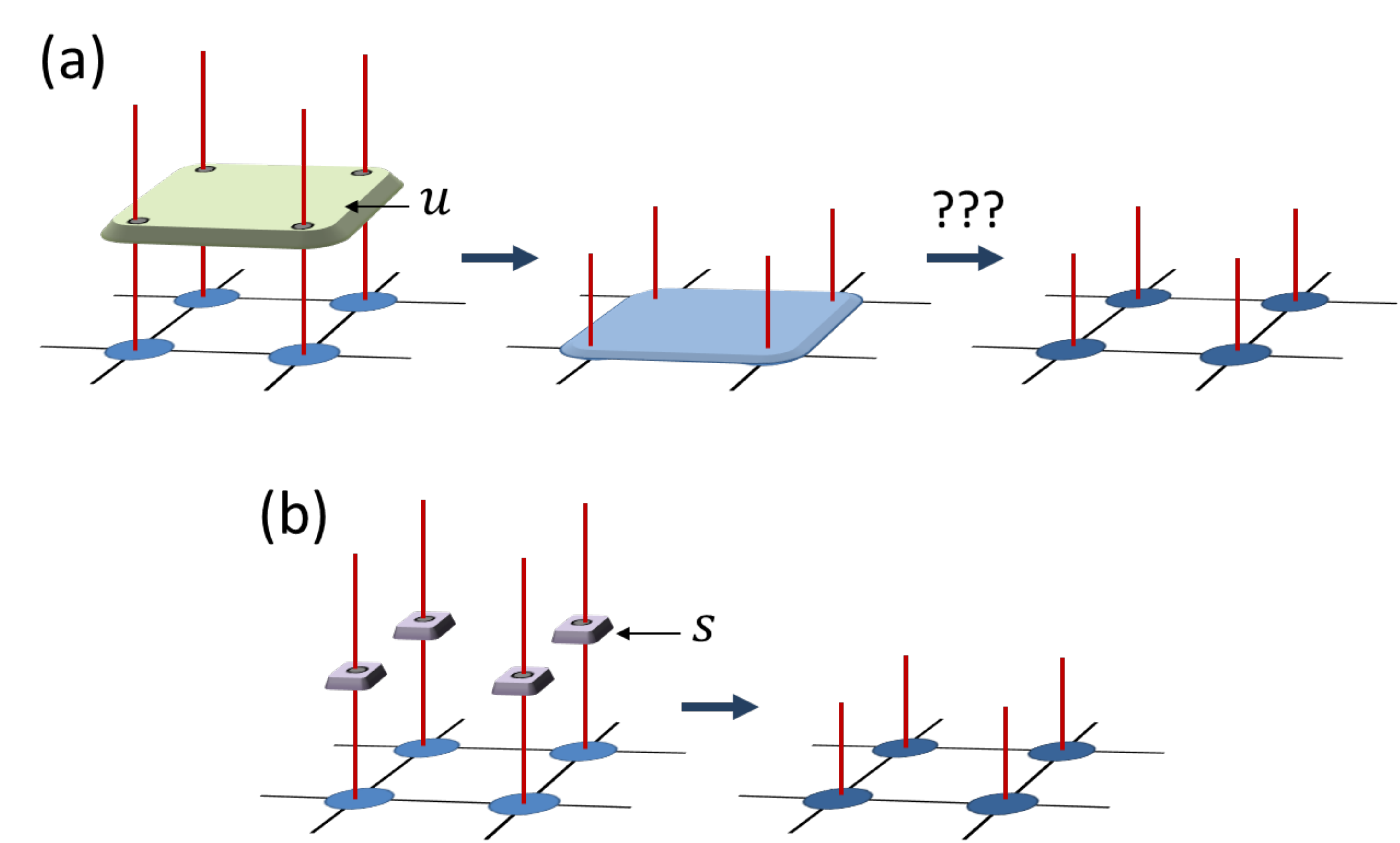}
\caption{(a) After applying a unitary disentangler $u$ to a $2\times 2$ plaquette from a PEPS, it is not clear how to recover the previous network structure. (b) By using implicit disentangling, the effect of $u$ could be replicated by a product of local operators $s$, which preserves the structure of the PEPS.}
\label{fig:CG2D}
\end{center}
\end{figure}

\section{Section D: Applications of implicit disentangling}
In this section we further detail some of the potential applications of IDR as a coarse-graining transformation for quantum states.

Already demonstrated in the main text is an algorithm for coarse-graining a quantum state described by a MPS, which could be straight-forwardly extended to higher spatial dimension for the coarse-graining of a PEPS \cite{PEPS1, PEPS2, PEPS3}. In this setting a major advantage of a coarse-graining transformation based on implicit disentangling over one based on unitary disentangling is apparent. If one were to try to employ ER, i.e. use unitary disentangling, to coarse-grain a PEPS, then it is not known how to recover the PEPS structure after applying a unitary disentangler $u$ to a $2\times 2$ plaquette, see Fig. \ref{fig:CG2D}(a). However, the PEPS structure is maintained by default when the equivalent disentangling is performed implicitly, as depicted in Fig. \ref{fig:CG2D}(b). We imagine than an algorithm for coarse-graining a PEPS using IDR could be useful in two different ways: (i) to evaluate data from an optimised PEPS, or (ii) to contract a PEPS as part of an optimization algorithm. In regards to the first application, this method could potentially allow more efficient evaluation of local expectation values, as the coarse-graining is applied to the PEPS state $\ket{\psi}$ directly, whereas previous methods involve contracting $\braket{\psi}{\psi}$ and thus squaring virtual dimension of the PEPS. It could be also applied to extract the conformal data of a PEPS that has been optimised for the ground state of a critical system, using the same approach as described in Sect. C of the supplementary material, which has not been demonstrated to be possible using any previous approach. In regards to the second application, we propose that IDR could be used to compute the local environments needed for the truncation step of iTEBD \cite{PEPS3}.   

\begin{figure}[!t!b]
\begin{center}
\includegraphics[width=8.6cm]{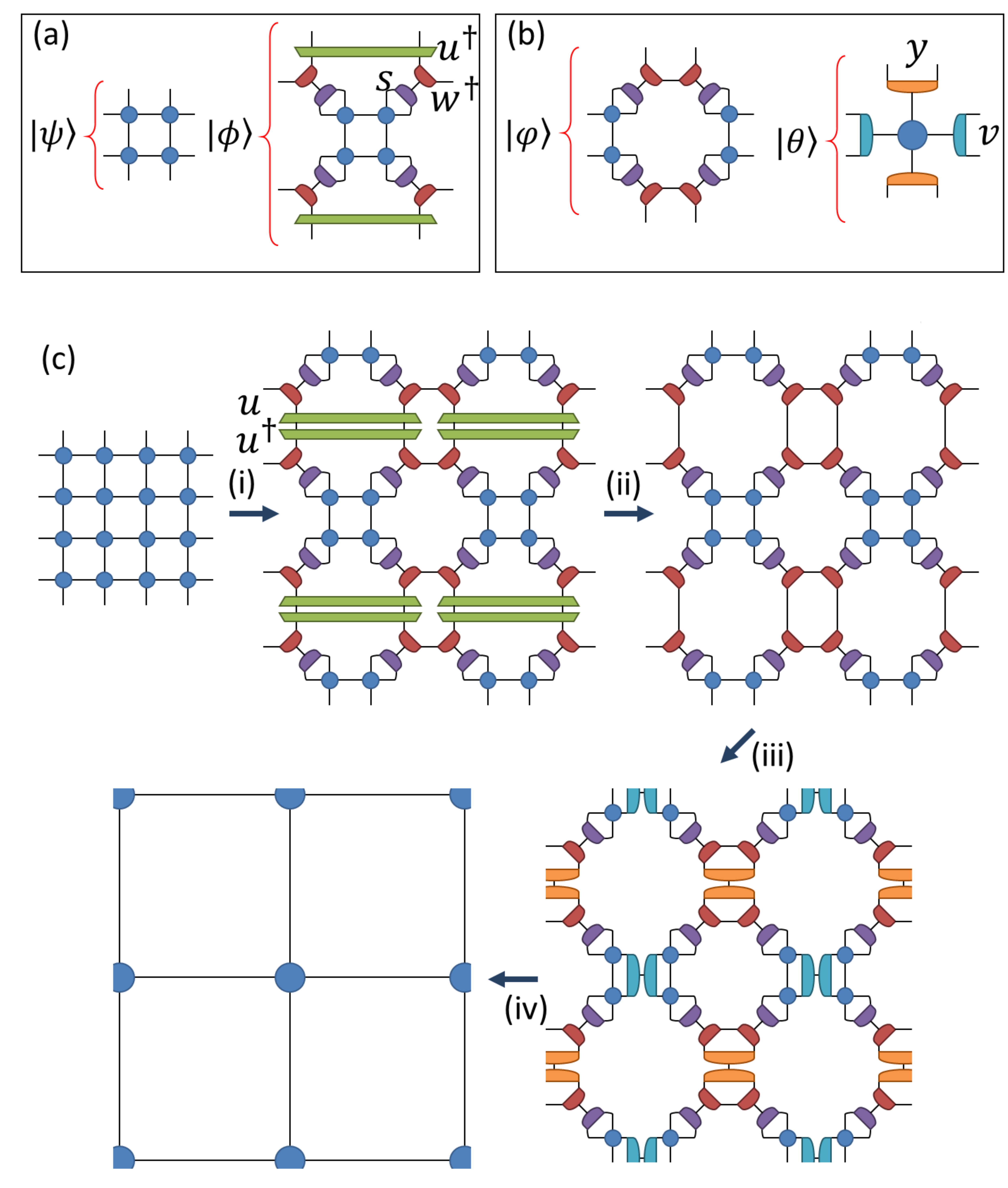}
\caption{An outline of a simplified TNR algorithm for a square-lattice tensor network. (a) Tensors $u$, $w$ and $s$ are optimised to maximise the fidelity between $\ket{\psi}$ and $\ket{\phi}$. (b) Isometries $y$ and $v$ are optimised to maximise the fidelity between $\ket{\varphi}$ and $\ket{\theta}$. (c-i) The first step of the TNR iteration is implemented by making the substitution of $\ket{\psi}$ with $\ket{\phi}$ for all $2\times 2$ plaquettes. (c-ii) The disentanglers $u$ cancel out with those from neighbouring cells, $u u^\dag = \mathbb I \otimes \mathbb I$, such that they are not used in the coarse-graining of the network. (c-iii) Cells of $\ket{\varphi}$ are substituted with those of $\ket{\theta}$. (c-iv) Tensors are contracted to form a new square-lattice network.}
\label{fig:TNR}
\end{center}
\end{figure}

Another potential application of IDR could be towards tensor network renormalization (TNR) \cite{TNR1, TNR2, TNR3, TNR4} algorithms. A sketch of how implicit disentangling can be incorporated into TNR for $2D$ tensor networks, which can be applied to study $2D$ classical systems or $1D$ quantum systems, is presented in Fig. \ref{fig:TNR}. Here the key difference from the standard TNR is that while unitary tensors $u$ are still optimised during the coarse-graining transformation, as in Fig. \ref{fig:CG2D}(a), they are not used in the actual coarse-graining of the tensor network, depicted in Fig. \ref{fig:CG2D}(c), as they cancel out. This has the benefit of not only simplifying the overall TNR algorithm, by allowing each iteration to be accomplished with fewer steps, but also of reducing the computational cost. The  tensor contractions in Fig. \ref{fig:CG2D}(c) can be accomplished with a cost that scales as $O(\chi^5)$ in the bond dimension $\chi$, as compared to the standard algorithm \cite{TNR4} which scales as $O(\chi^6)$. The optimization of the tensors necessary for the implicit disentangling, shown in Fig. \ref{fig:CG2D}(a), can also be accomplished in $O(\chi^5)$ cost by using similar tricks as explained in Ref. \cite{TNR4}, but we do not elaborate further here. It is important to note that the version of TNR based on implicit disentangling still produces a MERA (with proper isometric and unitary constraints) from a Euclidean path integral. As with the previous TNR algorithm \cite{TNR2}, when applied to coarse-grain a tensor network with an open boundary, the unitaries $u$ and isometries $y$ on the boundary no longer cancel and instead constitue a MERA. This contrasts with other recently proposed renormalization schemes for tensor networks \cite{TNR5, TNR6}, which achieve scale-invariance but do not produce MERA with unitary and isometric constraints.

The use of implicit disentangling is also likely to vastly improve efforts to implement the TNR approach for $3D$ tensor networks, which could be applied to study $3D$ classical or $2D$ quantum systems. In particular, if the action of a unitary disentangler applied to a $2\times 2$ plaquette could be replicated by local operators using implicit disentangling, similar to the scenario depicted in Fig. \ref{fig:CG2D}, then the cubic lattice structure of the network would be automatically preserved. Thus, if this disentangling technique were paired with the same blocking strategies previously employed in the higher-order tensor renormalization group (HOTRG) \cite{HOTRG1, HOTRG2, HOTRG3}, it may be possible to implement TNR for $3D$ lattices with only a modest increase in algorithmic and computational complexity over standard HOTRG. This remains an interesting avenue for future work.


\begin{thebibliography}{99}
\bibitem{FiniteCorr}
Since the initial MPS accurately describes the critical ground state only up to a finite length scale $\zeta$ the scale-invariance eventually breaks down (after approximately $z=10$ RG steps in this instance).
\end{thebibliography}
\end{document}